\documentclass[prd,aps,preprintnumbers,floatfix,showpacs,byrevtex,twocolumn]{revtex4}
\usepackage{amsmath}
\usepackage{graphicx}
\newcommand{\be}{\begin{equation}}
\newcommand{\ee}{\end{equation}}

\begin{document}
\preprint{
\vbox{
\hbox{ADP-08-05/T665}
}}

\title{Preconditioning Maximal Center Gauge with Stout Link Smearing in $SU(3)$}

\author{Alan~\'O~Cais$^1$}
\email{alan.ocais@adelaide.edu.au}

\author{Waseem~Kamleh$^1$}

\author{Kurt Langfeld$^2$}

\author{Ben~Lasscock$^1$}

\author{Derek~Leinweber$^1$}

\author{Peter~Moran$^1$}

\author{Andre~Sternbeck$^1$}

\author{Lorenz~von~Smekal$^1$}

\affiliation{$^1$Special Research Center for
the Subatomic Structure of Matter (CSSM), School of Chemistry \& Physics, 
University of Adelaide, SA 5005, Australia}

\thanks{We thank the Australian Partnership for Advanced Computing (APAC) and eResearch South Australia for generous grants of supercomputer time which have enabled this project. We also thank the University of Adelaide Faculty of Sciences for providing time on their Condor\texttrademark cycle-scavenging system. This work is supported by the Australian Research Council.}

\affiliation{$^2$School of Maths \& Stats, University of Plymouth,
Plymouth, PL4 8AA, England}

\date{July 2, 2008}

\begin{abstract}
Center vortices are studied in SU(3) gauge theory using Maximal Center Gauge (MCG) fixing. Stout link smearing and over-improved stout link smearing are used to construct a preconditioning gauge field transformation, applied to the original gauge field before fixing to MCG. We find that preconditioning successfully achieves higher gauge fixing maxima. We observe a reduction in the number of identified vortices when preconditioning is used, and also a reduction in the vortex-only string tension.
\end{abstract}

\vspace{3mm}

\maketitle

\newpage

\section{Introduction}

Despite more than 30 years of intense study, quark color confinement
in hadron physics remains unexplained (for a recent overview see
Ref.~\cite{Alkofer:2006fu}). Within the framework of lattice gauge
theory, the prevailing view is that quark confinement is the result of
a particular class of gauge field configurations which dominate the
QCD vacuum on large distance scales. Two potential candidates have
been most commonly investigated: confinement by means of $Z_N$ center
vortices and confinement due to Abelian monopoles (for a critical
discussion of both see Ref.~\cite{Greensite:2003bk}). To enhance these
particular features, gauge fields can be first fixed to a suitable
gauge, such as Maximal Abelian Gauge (MAG) \cite{Kronfeld:1987ri} or Maximal Center Gauge
(MCG) \cite{Del Debbio:1998uu}. Monopoles and center vortices are then defined by the
projection of these gauge-fixed fields onto $U(1)^{N-1}$ or $Z_N$,
respectively. Significant progress to date has occurred in $SU(2)$
using MAG and MCG, with original findings reproducing about 90\%
\cite{Bali:1996dm} and about 100\% \cite{Bertle:2000py}, respectively, of the
non-Abelian string tension. Removing monopole 
\cite{Miyamura:1995xn, Sasaki:1998ww, Bornyakov:2007fz} or center-vortex
\cite{de Forcrand:1999ms, Alexandrou:1999iy, Gattnar:2004gx, Gubarev:2005az, Bornyakov:2007fz, Bowman:2008qd}
 degrees of freedom from $SU(2)$ lattice gauge fields appears to leave
 topologically trivial, non-confining gauge fields that do not spontaneously break chiral symmetry.   

The significance of the center of the gauge group is what connects
possible candidates for this special class of configurations. As
outlined for the case of Laplacian Center Gauge (LCG) in
Ref.~\cite{deForcrand:2000pg},
 \emph{all} monopole world lines are embedded in 2-dimensional vortex surfaces. These topological objects naturally occur together as local gauge defects. In MCG it has been observed computationally that over 90\% of monopole currents are localized on center vortices \cite{Ambjorn:1999ym,Kovalenko:2004iu}. Strongly correlated effects between the two have also been observed by means of studying monopoles after vortex removal and vice-versa \cite{Boyko:2006ic}, as well as through the effect of their removal on the spectra of the overlap Dirac operator \cite{Bornyakov:2007fz}.

Again, all these advancements have been in $SU(2)$ and work in $SU(3)$
has not progressed to this level. While initial investigations were
hopeful \cite{Montero:1999by, Faber:1999sq}, subsequent results for
MAG and MCG were not so encouraging \cite{Stack:2002sy, Langfeld:2003ev, Leinweber:2006zq}, 
as they failed to reproduce the full non-Abelian 
string tension. Earlier studies in $SU(2)$ using MCG reported that
the center-projected configurations recovered the full
string-tension, however further study into the ambiguities of the
gauge-fixing procedure showed that this result is plagued by Gribov
copy effects \cite{Kovacs:1999st, Bornyakov:2000ig, Faber:2001hq}:
methods which give higher values of the gauge fixing functional
produce smaller values for the vortex-induced string tension. We
point out that when the Laplacian Center Gauge of
Refs.~\cite{Alexandrou:1999vx,deForcrand:2000pg}
 (which is free of Gribov ambiguities on the lattice) is used as the fixing method, the full $SU(3)$ (and $SU(2)$) string tension is recovered for the center-projected gauge fields but \emph{only} in the continuum limit. However, unlike MCG vortices \cite{Langfeld:1997jx}, the interpretation of LCG vortex matter is cumbersome in the same continuum limit \cite{Langfeld:2003ev, Langfeld:2001nz}.

In this paper we focus on the Gribov problem of the $SU(3)$ center-vortex picture of confinement using the MCG fixing method. We apply the ``smeared gauge fixing'' method of Ref.~\cite{Hetrick:1997yy} to MCG to ameliorate this Gribov problem. This creates a pre-conditioning gauge transformation for the configuration that should bring it closer to the global maximum. We investigate the effect of this method on the features of the long-distance behavior of the static quark potential as evaluated on configurations where the P-vortices derived from MCG have been removed and configurations composed purely of these P-vortices. In $SU(2)$, it has been shown that center-vortex removal specifically targets topological properties \cite{Gattnar:2004gx, Bornyakov:2007fz}, so as well as using stout-link smearing \cite{Morningstar:2003gk} we also employ over-improved stout-link smearing \cite{Moran:2008ra} to attempt to exploit the link to topological structure \cite{Ilgenfritz:2008ia}.

\section{Methodology}

\subsection{Identifying Vortex Matter}

In the center-vortex picture of confinement the gauge fields are considered to be decomposed into a long-range, smooth field $Z_\mu$ carrying all the confining fluctuations and a short-range field $V_\mu$ containing non-confining perturbations as well as other short-range effects
\begin{displaymath}
  U_\mu(x)=Z_\mu(x)V_\mu(x).
\end{displaymath}
Here $Z_\mu(x)$ is the center element which is closest, on the $SU(3)$ group manifold, to $U_\mu(x)$. A vortex is a configuration of the gauge potentials topologically characterized by non-trivial elements of $Z_3$ and is created by a singular gauge transformation. The non-trivial center element of the singular gauge-transformation characterizing the vortex may be made to be distributed over many links of an encircling loop (due to the short-range effects of $V_\mu(x)$). If we assume that by a gauge transformation the non-trivial center element can be concentrated on just one link we can compress this \emph{thick} vortex into a \emph{thin} one. If we then project this gauge transformed configuration onto its center elements, the projected vortices (P-vortices) linking with the loop should then correspond to the thin vortex. It is for this reason that we adopt the use of gauge-fixing to obtain the necessary gauge transformation. It is the choice of gauge that determines our method for finding the center vortices and, therefore, the connection between the P-vortices and the thick center vortices present in the original configuration. The particular choice of gauge, the properties of the P-vortices associated with each choice and the Gribov problem that it creates is what has polarized opinions in this area \cite{Faber:1998en,Kovacs:1999st, Faber:1999hw, deForcrand:2000pg, Langfeld:2003ev}.

Here, we employ the MCG gauge-fixing algorithm as outlined in Ref.~\cite{Langfeld:2003ev}. The gauge condition we chose to maximize (with respect to the center elements $Z_\mu(x)$) in this algorithm is 
\begin{displaymath}
V_U[\Omega] = \frac{1}{N_l}\sum_{x,\mu}\big[\frac{1}{3}\text{tr}U^{\Omega}_\mu(x)\big]\big[\frac{1}{3}\text{tr}U^{\Omega}_\mu(x)\big]^\dagger,
\end{displaymath} 
where $N_l$ is the number of links on the lattice and $U^{\Omega}$ is the gauge-transformed field.

After fixing the gauge, each link should be close to a center element of $SU(3)$, $Z^m=e^{i\phi^m}$, $\phi^m=\frac{2\pi}{3}m$ with $m\in \{ -1,0,1 \}$.
Since, for every link, 
\begin{displaymath}
\frac{1}{3}\text{tr} U^{\Omega}_\mu(x) = u_{x,\mu}e^{i\phi_{x,\mu}} \hspace{0.3cm}\text{and}\hspace{0.3cm} \phi_{x,\mu}=\tan^{-1}\frac{\text{Im}(\text{tr} U^{\Omega}_\mu(x))}{\text{Re}(\text{tr} U^{\Omega}_\mu(x))}
\end{displaymath}
then $\phi_{x,\mu}$ should be near to some $\phi^m$, by construction of the gauge-fixing condition. We then perform the center projection by mapping
\begin{displaymath}
SU(N)\mapsto Z_N :\hspace{0.2cm} U^{\Omega}_\mu(x)\mapsto Z_\mu(x) \hspace{0.3cm} \text{with}\hspace{0.3cm} Z_\mu(x)=e^{i\phi^m_{x,\mu}},
\end{displaymath}
with the appropriate choice of $\phi^m_{x,\mu}$, $m\in \{ -1,0,1 \} $.

To reveal the vortex matter we simply take a product of links around an elementary plaquette. We say a vortex pierces the plaquette if this product is a non-trivial center element and the plaquette is then a \emph{P-vortex}. We can remove these P-vortices by hand from the configuration using $U_\mu^\prime(x)=Z^\dagger_\mu(x)U_\mu^\Omega(x)$.

\subsection{Smearing as a Preconditioner}

In the center-vortex picture of confinement, the center elements correspond to the long-range physics. It would seem reasonable then to employ the use of smearing to smooth out the short-range fluctuations and allow the gauge transformation to see more of the underlying long-range physics. To this end we construct a preconditioning gauge transformation for each gauge field to obtain higher maxima in the gauge-fixing procedure and thereby directly address the Gribov-copy issue \cite{Hetrick:1997yy}.

Firstly, we smear the gauge field using any smearing algorithm (stout-link smearing \cite{Morningstar:2003gk} has been applied here as well as over-improved stout-link smearing which has been shown to better preserve the topological structure underlying the original configuration \cite{Moran:2008ra}). We then fix the smeared field using the MCG gauge-fixing method. At each iteration we keep track of the total gauge transformation that has been applied to the smeared gauge field. Once the algorithm has converged we use the stored total transformation as a preconditioning gauge transformation for the unsmeared gauge field. We emphasize that the (unsmeared) preconditioned gauge field remains on the same gauge orbit since the preconditioning is merely a (specific) gauge transformation on the original links. Gauge-fixing the preconditioned field simply gives us a Gribov-copy of the result from gauge-fixing the original gauge field.

\section{Results}
Calculations are performed using 200 quenched configurations with the L\"uscher-Weisz plaquette plus rectangle gauge action \cite{Luscher:1984xn} on a $20^3\times40$ lattice with $\beta=4.52$. Similar preliminary results have being found on 100 $16^3\times32$ lattices (with $\beta=4.6$) and have been reported elsewhere \cite{Cais:2007bm}.

Stout-link smearing with a smearing parameter of $0.1$ is used to construct the preconditioning transformation with the number of sweeps ranging from 0 to 20 in steps of 4 sweeps. We also employ over-improved stout-link smearing with a smearing parameter of 0.06 and an $\epsilon$ parameter of $-0.25$. Here, each preconditioning was conducted independently.

\begin{table}
  \begin{tabular}{c@{\hspace{0.4cm}} c@{\hspace{0.4cm}} c@{\hspace{0.4cm}} c@{\hspace{0.4cm}} c}
    \hline
    \hline
    \textbf{Sweeps} & \textbf{Iteration} & \textbf{Smear} & \textbf{Max}& \textbf{Vortices}\\
     & \textbf{Blocks} & \textbf{Max} & & \\
    \hline
    0&$80\pm20$&$-$&0.7350(7)& $ 3.21(12)\% $ \\
    4&$118\pm22$&0.9150(11)&0.7400(6)& $ 1.93(10)\% $ \\
    8&$126\pm26$&0.9369(54)&0.7407(6)& $ 1.71(10)\% $ \\
    12&$126\pm21$&0.9459(12)&0.7411(6)& $ 1.58(10)\% $ \\
    16&$128\pm23$&0.9506(12)&0.7412(6)& $ 1.53(10)\% $ \\
    20&$135\pm26$&0.9541(12)&0.7414(5)& $ 1.45(11)\% $ \\
    OI 80&$148\pm29$&0.9625(14)&0.7417(6)& $ 1.28(13)\% $ \\
    \hline
    \hline
  \end{tabular}
  \caption{\textnormal{Results for the average maximum gauge condition $V_U[\Omega]$ as a function of preconditioning stout-link smearing sweeps (OI signifies over-improved stout-link smearing). For each of the sweeps used in the preconditioning: the average total (smeared gauge field fixing plus preconditioned gauge field fixing) number of blocks of 50 iterations used, the average smeared gauge condition maximum reached, the average preconditioned gauge condition maximum reached and the percentage of plaquettes that are P-vortices.}}
  \label{tab:errors} 
\end{table}

Given that the original goal was to increase the gauge-fixing maxima achieved in MCG fixing, we can see from Table~\ref{tab:errors} that we are successful, in this regard, in every case. With each level of preconditioning a higher gauge condition maximum is achieved both for the smeared gauge field and the preconditioned original field. If we compare $0$ and $4$ sweeps of preconditioning, we can see that the magnitude of this increase is initially large but the increase is slower as we precondition to higher levels. However this increase does not come without a cost, the number of gauge-fixing iteration blocks (a block is 50 iterations) required almost doubles between the unpreconditioned fixing and the maximum amount of preconditioning. Typically, two-thirds of the iterations are spent fixing the smeared field and one-third fixing the preconditioned field. 

What is most significant about this table however is that with each level of preconditioning the percentage of projected plaquettes which are P-vortices drops significantly. Without preconditioning $3.21\%$ of plaquettes are vortices and this drops to as low as $1.28\%$ for the highest level of preconditioning.

\begin{table}
  \centering
  \begin{tabular}{c| c@{\hspace{0.4cm}} c@{\hspace{0.4cm}} c@{\hspace{0.4cm}} c@{\hspace{0.4cm}} c@{\hspace{0.4cm}} c@{\hspace{0.4cm}} c}
    \hline
    \hline
    \textbf{Sweeps} & \textbf{0} & \textbf{4} & \textbf{8} & \textbf{12} & \textbf{16} & \textbf{20} &  \textbf{OI 80}\\
    \hline
    \textbf{0} & $\ddots$ & 100 & 100 & 100 & 100 & 100 & 100 \\
    \textbf{4} & $39 ^{\pm 4}$ & $\ddots$ & 96.5 & 100 & 100 & 100 & 100 \\
    \textbf{8} & $46 ^{\pm 4}$ & $11 ^{\pm 6}$ & $\ddots$ & 81.5 & 91.5 & 97.5 & 99.5 \\
    \textbf{12} & $50 ^{\pm 4}$ & $17 ^{\pm 6}$ & $ 9 ^{\pm 6}$ & $\ddots$ & 69 & 81 & 97.5 \\
    \textbf{16} & $52 ^{\pm 3}$ & $20 ^{\pm 6}$ & $12 ^{\pm 6}$ & $7 ^{\pm 5}$ & $\ddots$ & 69.5 & 95.5 \\
    \textbf{20} & $54 ^{\pm 4}$ & $24 ^{\pm 7}$ & $15 ^{\pm 7}$ & $10 ^{\pm 7}$ & $8 ^{\pm 6}$ & $\ddots$ & 85.5 \\
    \textbf{OI 80} & $59 ^{\pm 4}$ & $33 ^{\pm 7}$ & $24 ^{\pm 8}$ & $19 ^{\pm 9}$ & $16 ^{\pm 9}$ & $14 ^{\pm 8}$ & $\ddots$\\
    \hline
    \hline
  \end{tabular}
  \caption{\textnormal{Comparisons between different preconditioning levels of stout-link smearing (OI signifies over-improved stout-link smearing). The upper triangle of this table (from preconditioning level row to preconditioning level column) we report the percentage of configurations that experience a reduction in the measured number of P-vortices. The lower triangle (preconditioning level column to preconditioning level row) of this table gives the percentage reduction of the number of P-vortices for the configurations that experienced a reduction.}}
  \label{tab:transitions_vort_dec}
\end{table}

In Table~\ref{tab:transitions_vort_dec} we investigate further by looking at this particular effect between all the different levels of preconditioning. In the upper triangle of this table (from preconditioning level row to preconditioning level column) we report the percentage of configurations that experience a reduction in the measured number of P-vortices. As we can see, this percentage is always high but the effect is lessened as we move to transitions, particularly small transitions, between higher levels of preconditioning. It should be noted however that the relative difference between, for example, 20 sweeps of stout-link and 80 sweeps of over-improved stout-link preconditioning is difficult to quantify but the effect is still significant for this transition. 

The magnitude of this effect is also reported in Table~\ref{tab:transitions_vort_dec}. When reading from the lower triangle (preconditioning level column to preconditioning level row) of this table we can see the percentage reduction of the number of P-vortices for the configurations that experienced a reduction. In the transition from no preconditioning to any other level, the order of a 50\% reduction is observed. For other transitions it would seem the effect drops to the 10\% level reasonably quickly, but again we see an increased effect when we consider over-improved smearing. It should be noted that, regardless of the preconditioning level, the center phases of the links of the fields always remain evenly distributed across the three possible values, reflecting the fact that the realization of center symmetry remains unaffected.

\begin{table}
  \centering
  \begin{tabular}{c|c@{\hspace{0.4cm}} c@{\hspace{0.4cm}} c@{\hspace{0.4cm}} c@{\hspace{0.4cm}} c@{\hspace{0.4cm}} c@{\hspace{0.4cm}} c}
   \hline
   \hline
   \textbf{Sweeps} & \textbf{0} & \textbf{4} & \textbf{8} & \textbf{12} & \textbf{16} & \textbf{20} &  \textbf{OI 80}\\
   \hline
   \textbf{0} & $\ddots$ & 100 & 100 & 100 & 100 & 100 & 100 \\
   \textbf{4} & 100 & $\ddots$ & 89.5 & 98 & 99 & 100 & 100 \\
   \textbf{8} & 100 & 100 & $\ddots$ & 75 & 78.5 & 89.5 & 93.5 \\
   \textbf{12} & 100 & 100 & 95.33 & $\ddots$ & 59 & 71.5 & 87.5 \\
   \textbf{16} & 100 & 100  & 98.09 & 97.46 & $\ddots$ & 62.5 & 77.5 \\
   \textbf{20} & 100 & 100 & 100 & 97.2 & 93.6 & $\ddots$ & 71 \\
   \textbf{OI 80} & 100 & 100 & 100 & 100 & 100 & 97.18 & $\ddots$\\
   \hline
   \hline
  \end{tabular}
  \caption{\textnormal{Comparisons between different preconditioning levels of stout-link smearing (OI signifies over-improved stout-link smearing). When reading from sweep row to sweep column (upper triangle) the value shown is the percentage of configurations that achieve a higher gauge fixing maximum. When reading from sweep column to sweep row (lower triangle) the value shown is the percentage of the configurations with higher maximum that achieve a lower number of P-vortices.}}
  \label{tab:transitions_vort_max}
\end{table}

We can look to Table~\ref{tab:transitions_vort_max} when considering whether a higher gauge-fixing maximum translates into a lower number of P-vortices. When reading from preconditioning level row to preconditioning level column, the percentage of configurations that experience an increase in the gauge-fixing maximum is shown. Similar trends to that of Table~\ref{tab:transitions_vort_dec} are observed, with large effects initially which become reduced for small transitions between higher levels. Of these configurations with an increased maximum we can see  almost exclusively (when reading from preconditioning level column to preconditioning level row) that an increased gauge-fixing maximum does lead to a lower number of P-vortices. 

\begin{figure*}[h]
\hspace{-1cm}
$\begin{array}{c@{\hspace{1cm}}c}
  \mbox{\bf No Smearing Preconditioning} & \mbox{\bf 80 Sweeps OverImp. Preconditioning}\\ 
  \includegraphics[height=5.4cm]{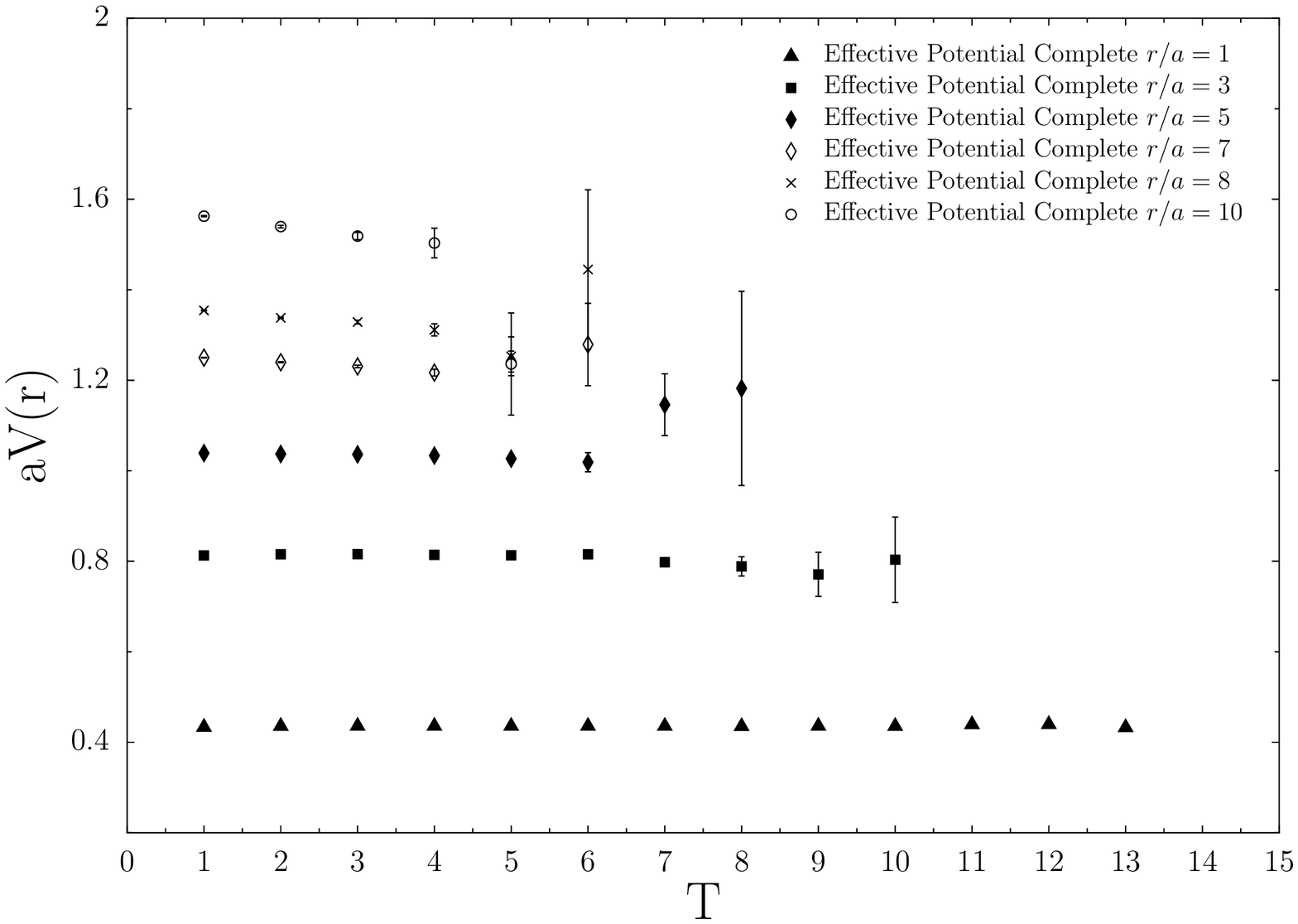} &
  \includegraphics[height=5.4cm]{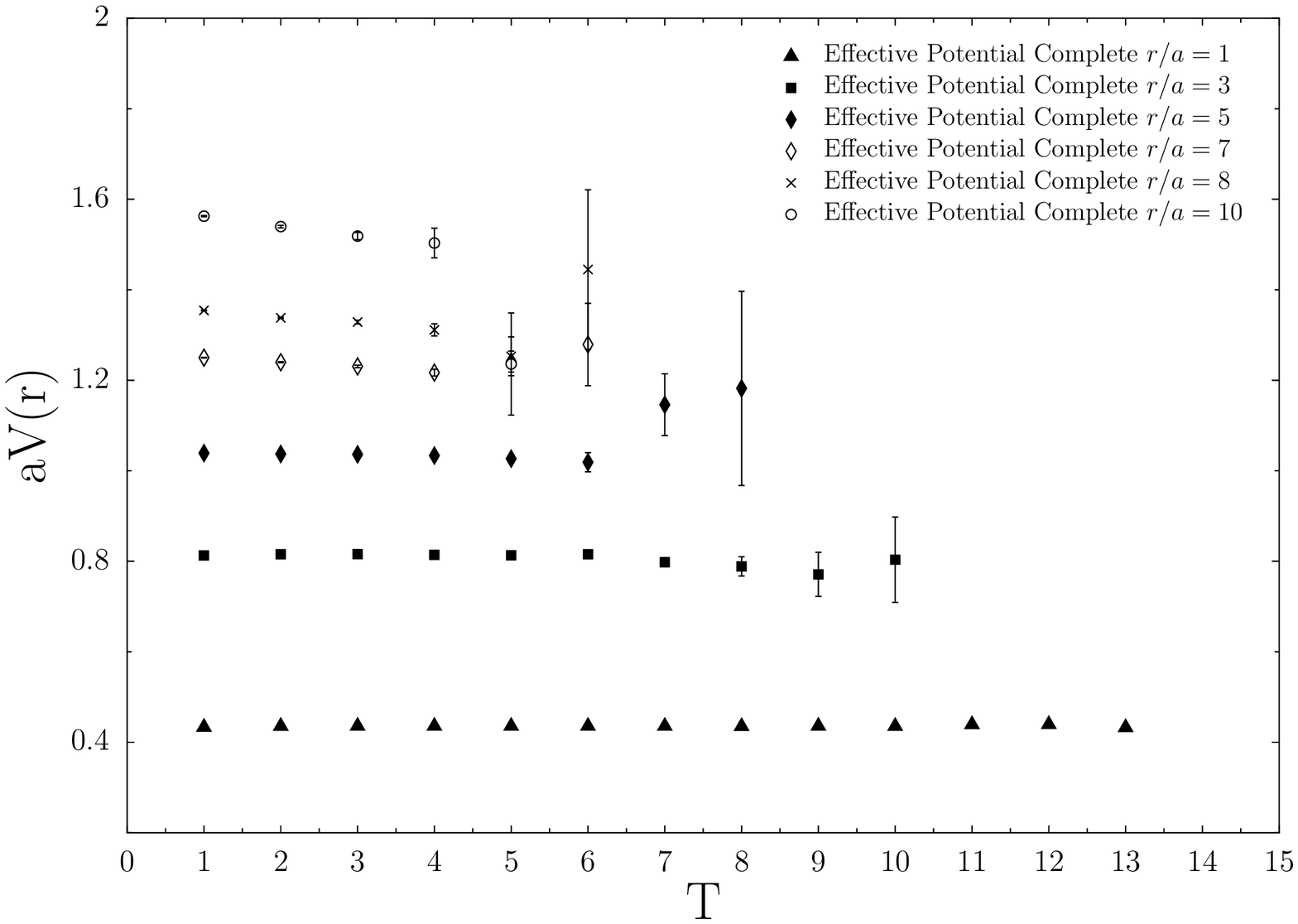} \\ [0.4cm] 
  \includegraphics[height=5.4cm]{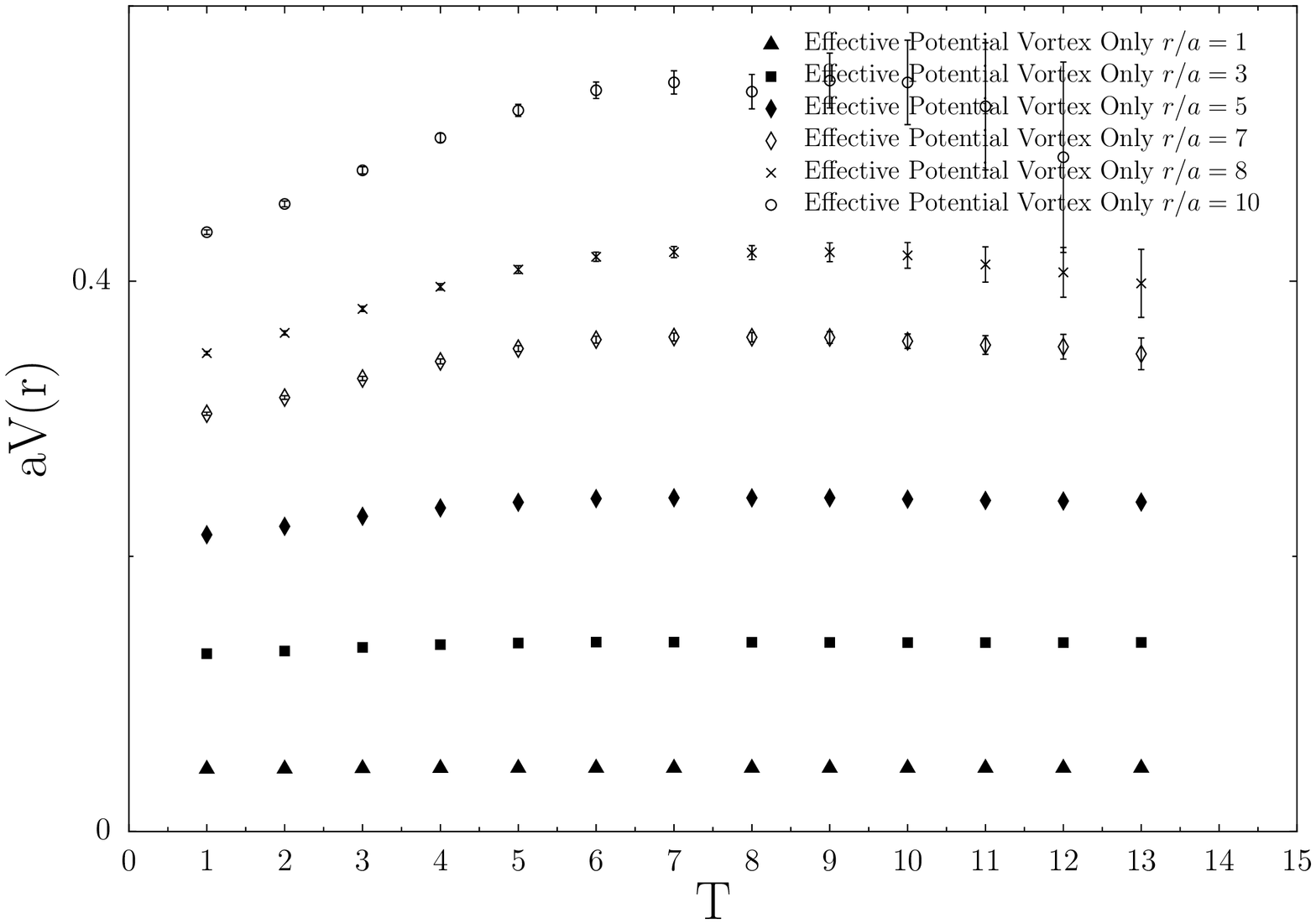}&
  \includegraphics[height=5.4cm]{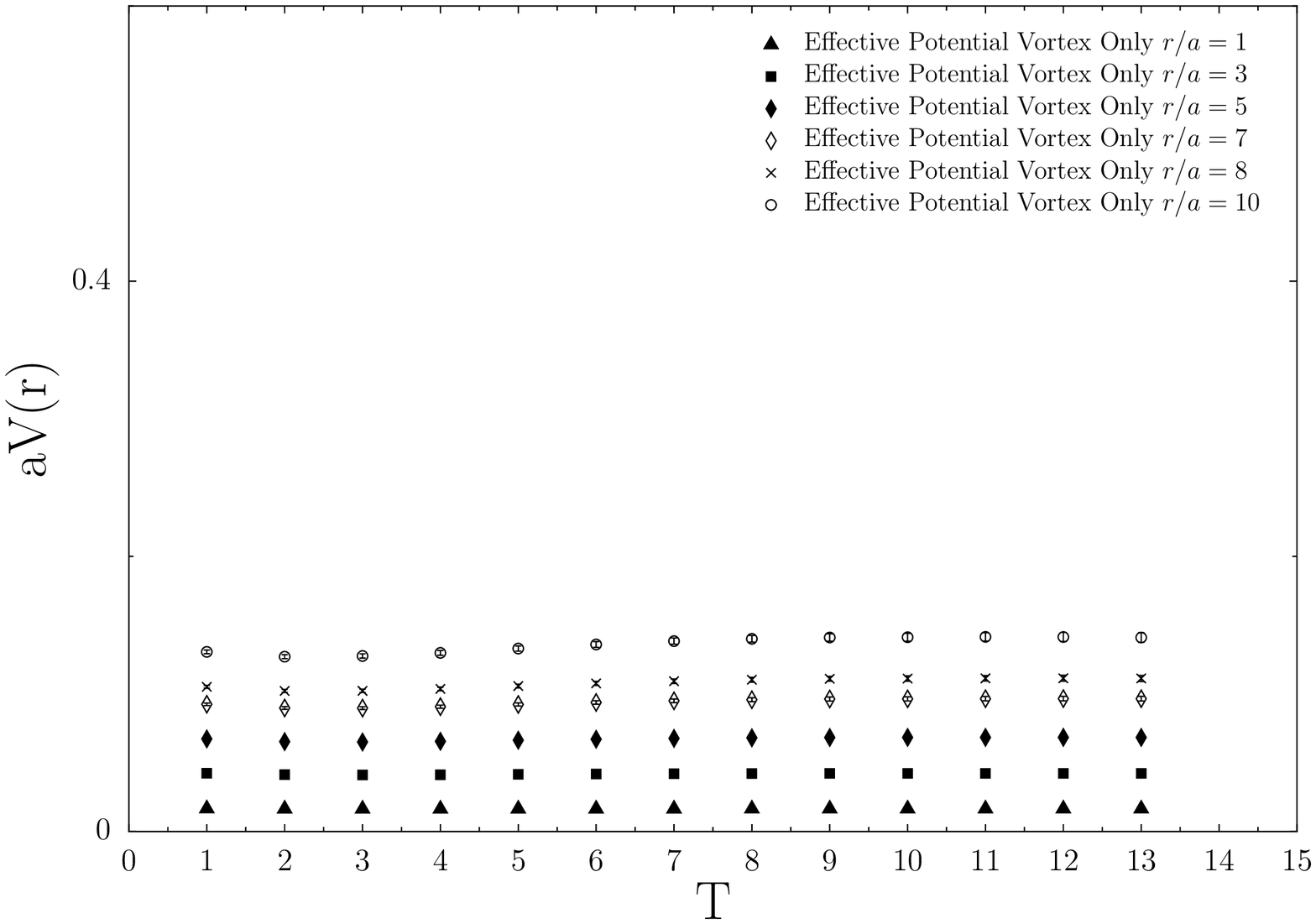}\\ [0.4cm] 
  \includegraphics[height=5.4cm]{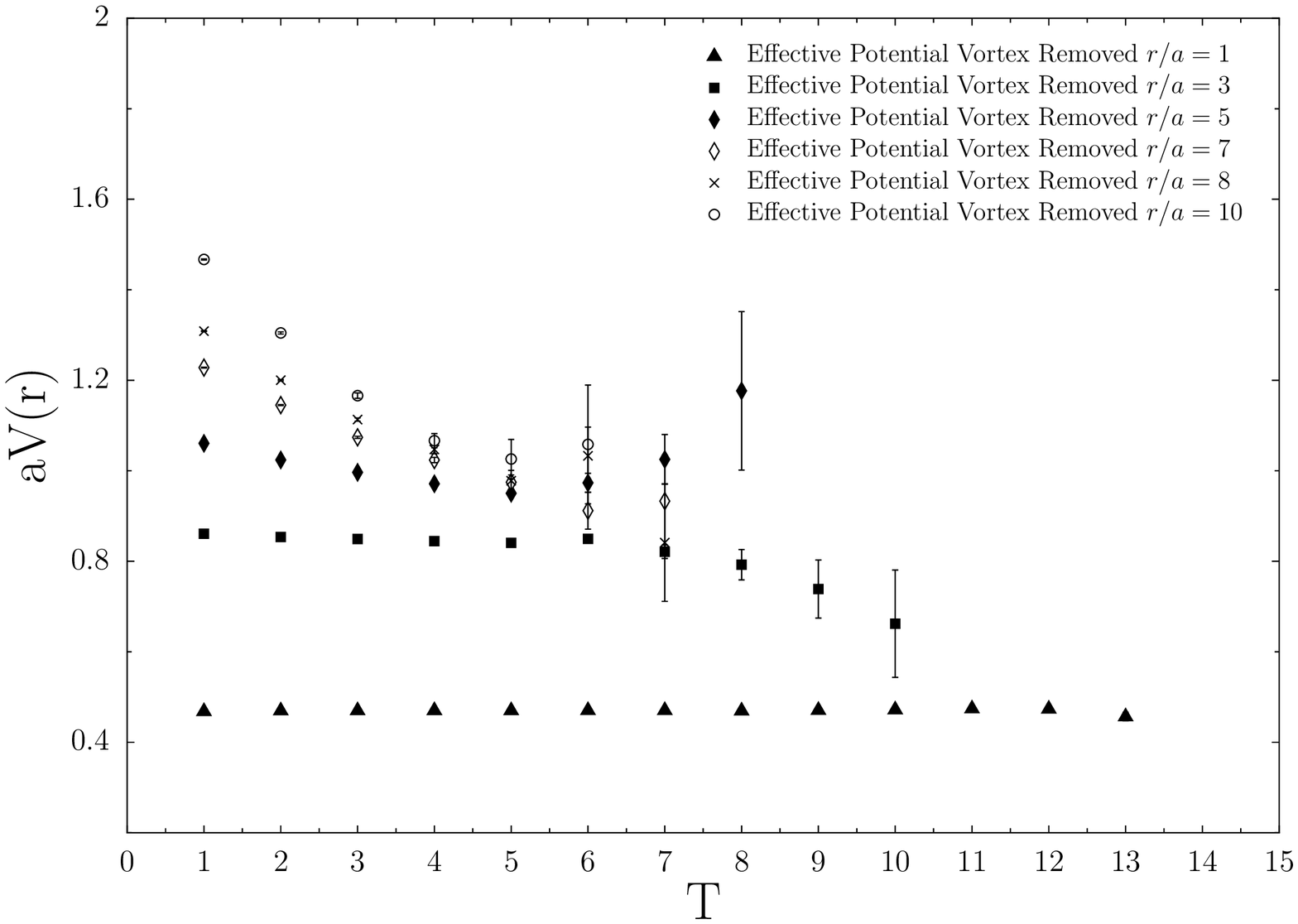} &
  \includegraphics[height=5.4cm]{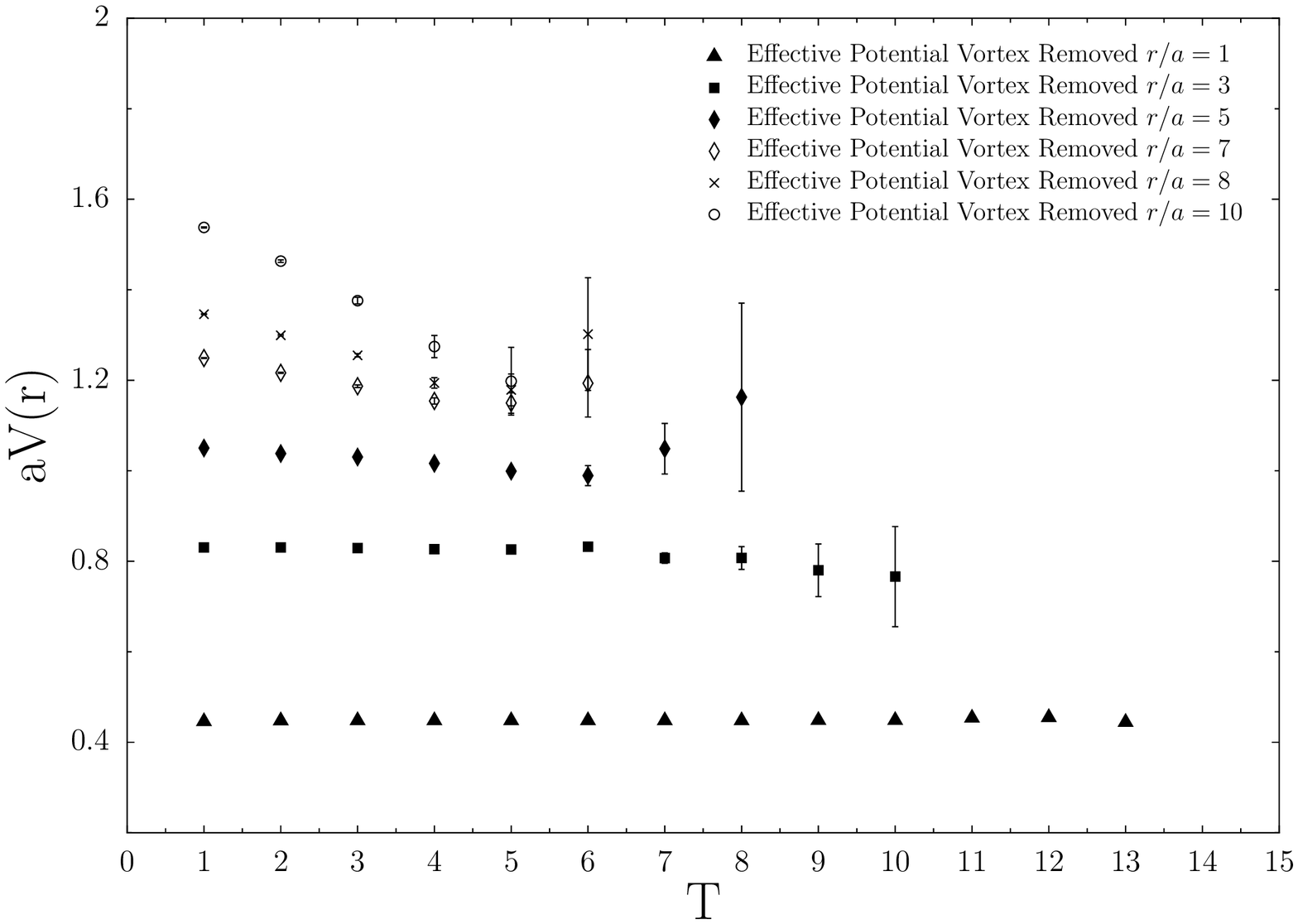}
  
\end{array}$
\caption{The effective potential plots  for the lowest (left) and highest (right) levels of preconditioning smearing. The upper plots contain the data for the original gauge-fixed configurations, the middle plots contain the data for the vortex-only configurations and the lower plots contain the data for the vortex-removed configurations. Each plot contains data for a range of quark separations.}
\label{figEffms}
\end{figure*}

\subsection{The Static Quark Anti-quark Potential}

The fact that we can reduce the number of P-vortices through preconditioning is not necessarily a cause for concern. As stated previously, our method for determining the location of center vortices is only justified by the physical relevance of the P-vortices that we determine. A first step in determining this relevance is the calculation of the static quark anti-quark potential. In the center-vortex picture, the string tension $\sigma$ as determined from the infrared behavior of this potential should be fully accounted for by the center-vortex component of the gauge fields, $Z_\mu$, with the Coulombic term accounted for by the vortex removed component, $V_\mu$. Since we can ``remove'' the determined P-vortices by the operation $U_\mu^\prime(x)=Z^\dagger_\mu(x)U_\mu^\Omega(x)$, we can seek to observe these properties directly. However, since the determined P-vortices are gauge-dependent (and their number Gribov-copy dependent, as we have already seen) then so too are the subsequent measurements of the static quark potential from the vortex-only and vortex-removed components of the configuration.

Computing the static quark anti-quark potential as a function of the quark separation is a two step process. Wilson loops $W(R,T)$ of extension of $R\times T$ have the large $T$ behavior
\begin{displaymath}
\langle W(R,T)\rangle \propto \text{exp}\{ -V(r)aT\},\hspace{1cm} r:=Ra,
\end{displaymath}
 where $a$ is the lattice spacing. The method for extracting the effective potential is thus identical to that of extracting effective masses from two-point functions in hadronic spectroscopy. To obtain the static quark anti-quark potential as a function of the quark separation we simply repeat this process for a range of values of the separation $R$. By using off-axis spatial paths for the Wilson loops, we can obtain non-integer values of $R$. We exploit full space-time translation to improve the statistics of our Wilson loops.

Since the final plot is composed of fits performed on a large number of effective potential plots for all the different separations, it is prudent and necessary that the factors determining those fits are given, and taken into account, when analyzing the subsequent static quark anti-quark potential as a function of separation. The difficulties associated with such fits can be easily recognized in Fig.~\ref{figEffms}. In these plots we show the static quark potential for a variety of quark separations for each of the original, vortex-only and vortex-removed configurations. On the left we show these plots for the unpreconditioned MCG fixing and on the right we show the same plots for 80 sweeps of over-improved smearing as a preconditioner.

One of the first things to discuss is the difficulty in obtaining a satisfactory fit range for the data, particularly in the case of the vortex-removed configurations. With these configurations, more so at larger separations, the potential falls rapidly and decays into noise quickly. A visually satisfying plateau region is not evident and we must rely on the fitting routine to determine the goodness of the fit. What the plot can tell us is that the effective potential continues to fall (at separations of 5 lattice spacings and greater) until at least time-slice 5. Since the data decays into noise around this point, we chose to constrain our fit using timeslice 5 and fit from this slice to slice 7 (a straight-line fit to 3 points). This constraint is then applied to all values of the separation. What we find is that while this may lead to reasonable ($\lesssim 1.3$) values of the $\chi^2$ per degree of freedom in the majority of cases, there are certainly significant deviations from this desirable result.

Global fit ranges are chosen in a somewhat similar way for the unpreconditioned and vortex-only configurations. For the unpreconditioned configurations, the global fit-range was chosen to be between timeslice 4 and 6 since these accounted for the systematic drift of the potential at large separations for lower time values and also gave reasonable $\chi^2$ behavior. In the case of the vortex-only configurations, the errors are far more controlled but the potential rises at small times and plateaus far later so the fit range was chosen to be from timeslice 10 to 12 but again some of the $\chi^2$ per degree of freedom values were unsatisfactory. This is most likely due to the heavy constraints placed on the fit by the accurate potential determinations.

\begin{figure*}[h]
\hspace{-1cm}
$\begin{array}{c@{\hspace{1cm}}c}
  \mbox{\bf No Smearing Preconditioning} & \mbox{\bf 4 Sweeps Stout Preconditioning}\\ 
  \includegraphics[height=5.4cm]{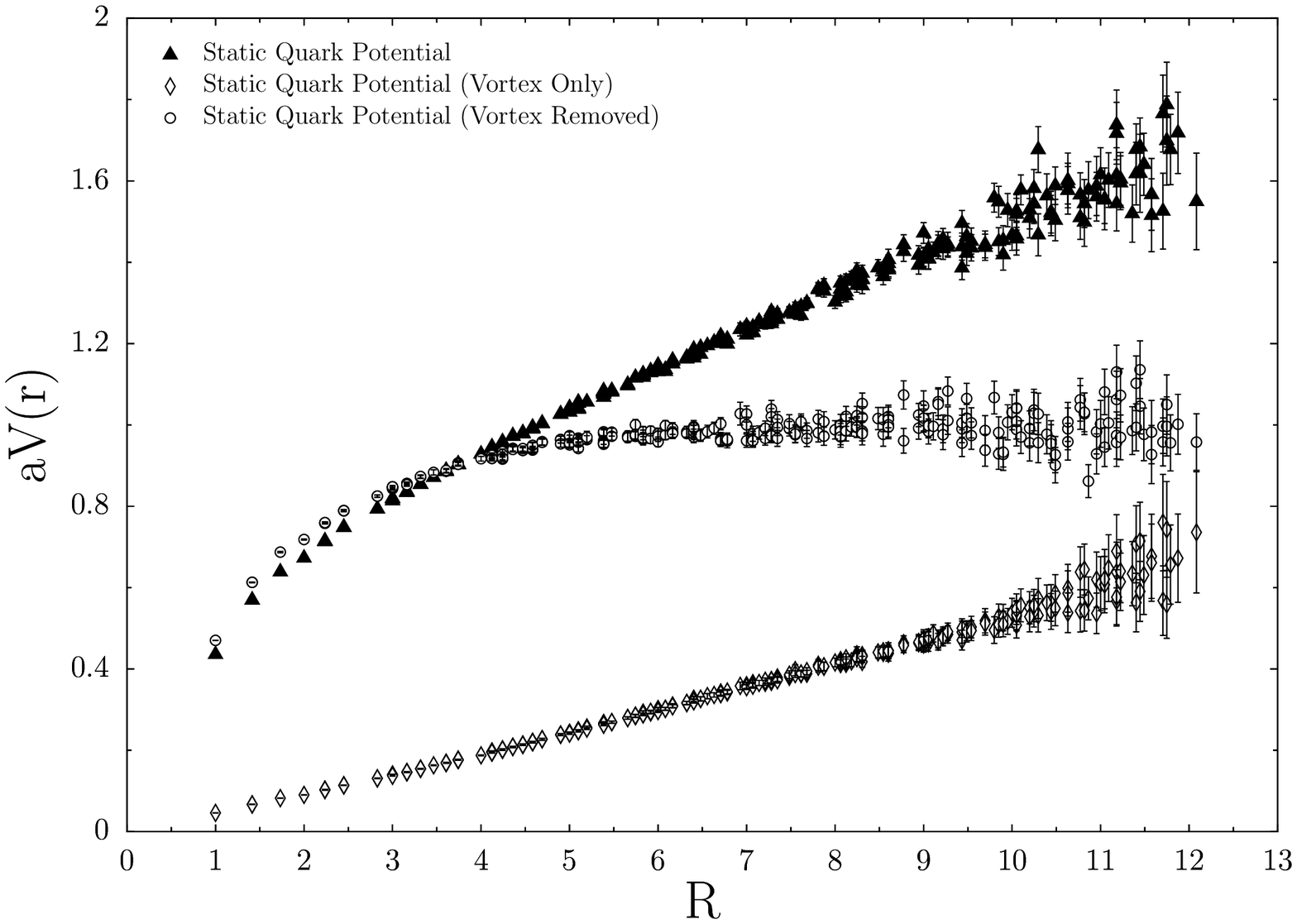} &
  \includegraphics[height=5.4cm]{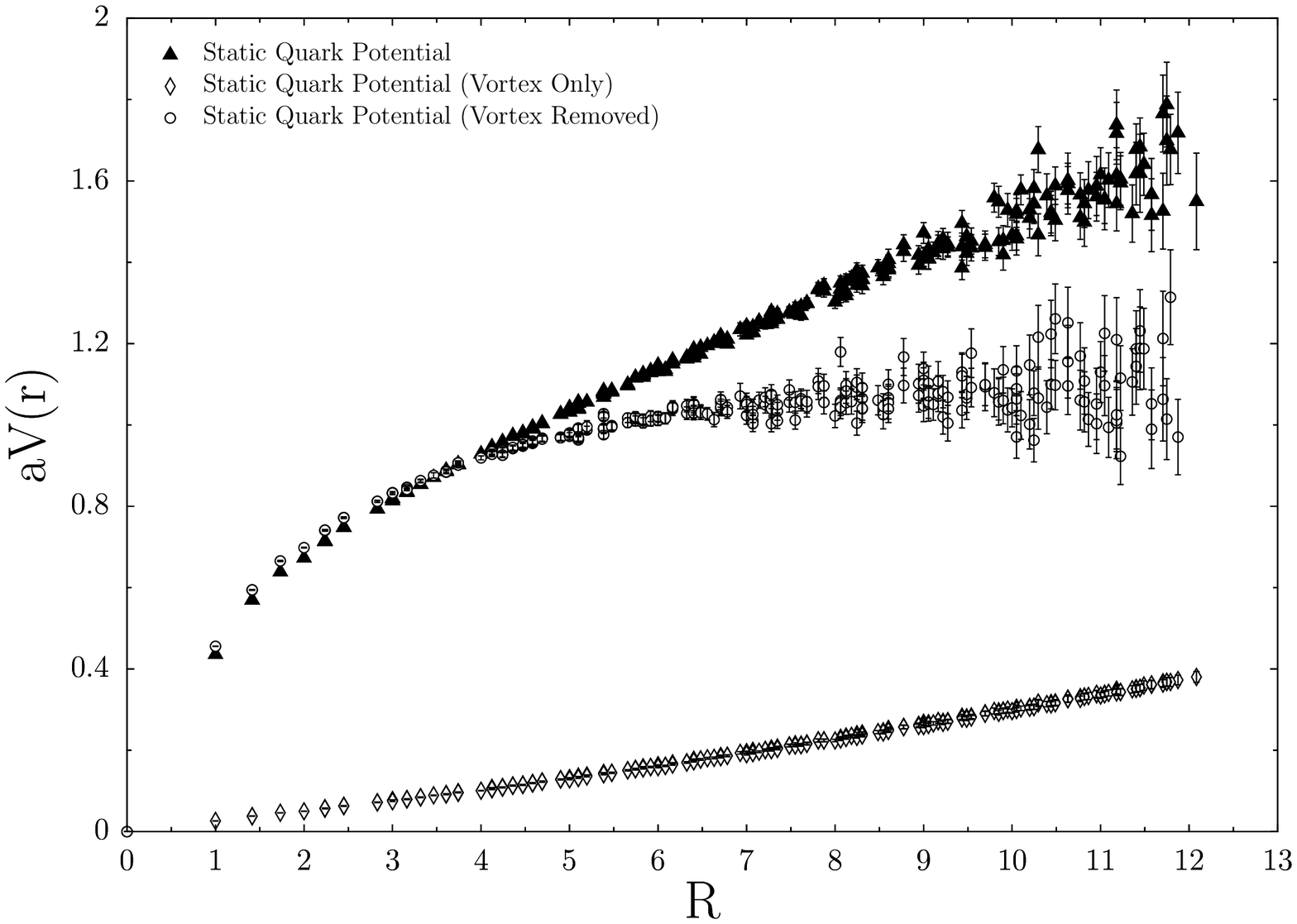} \\ [0.4cm] 
  \mbox{\bf 16 Sweeps Stout Preconditioning} & \mbox{\bf 80 Sweeps OverImp. Preconditioning}\\ 
  \includegraphics[height=5.4cm]{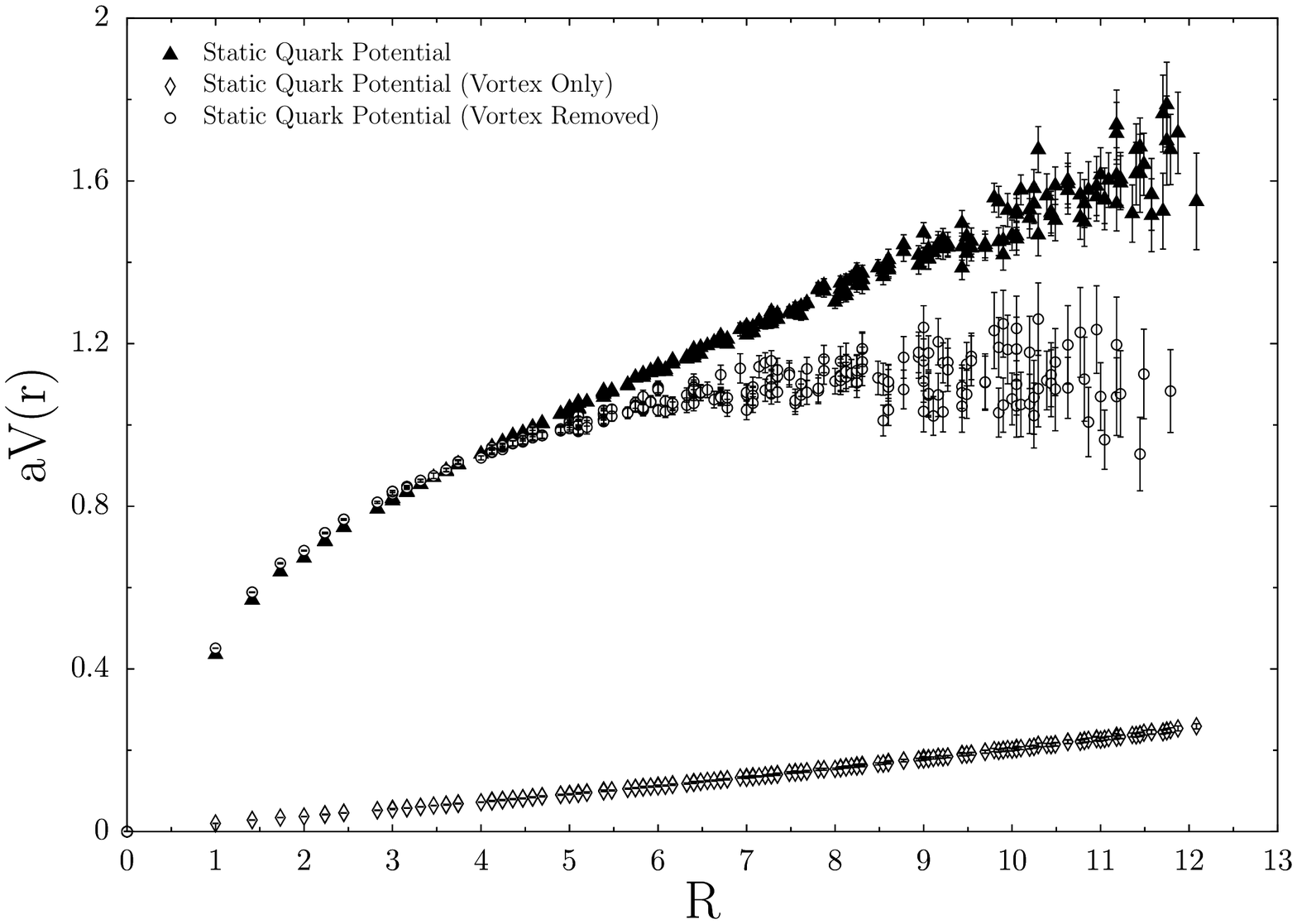} &
  \includegraphics[height=5.4cm]{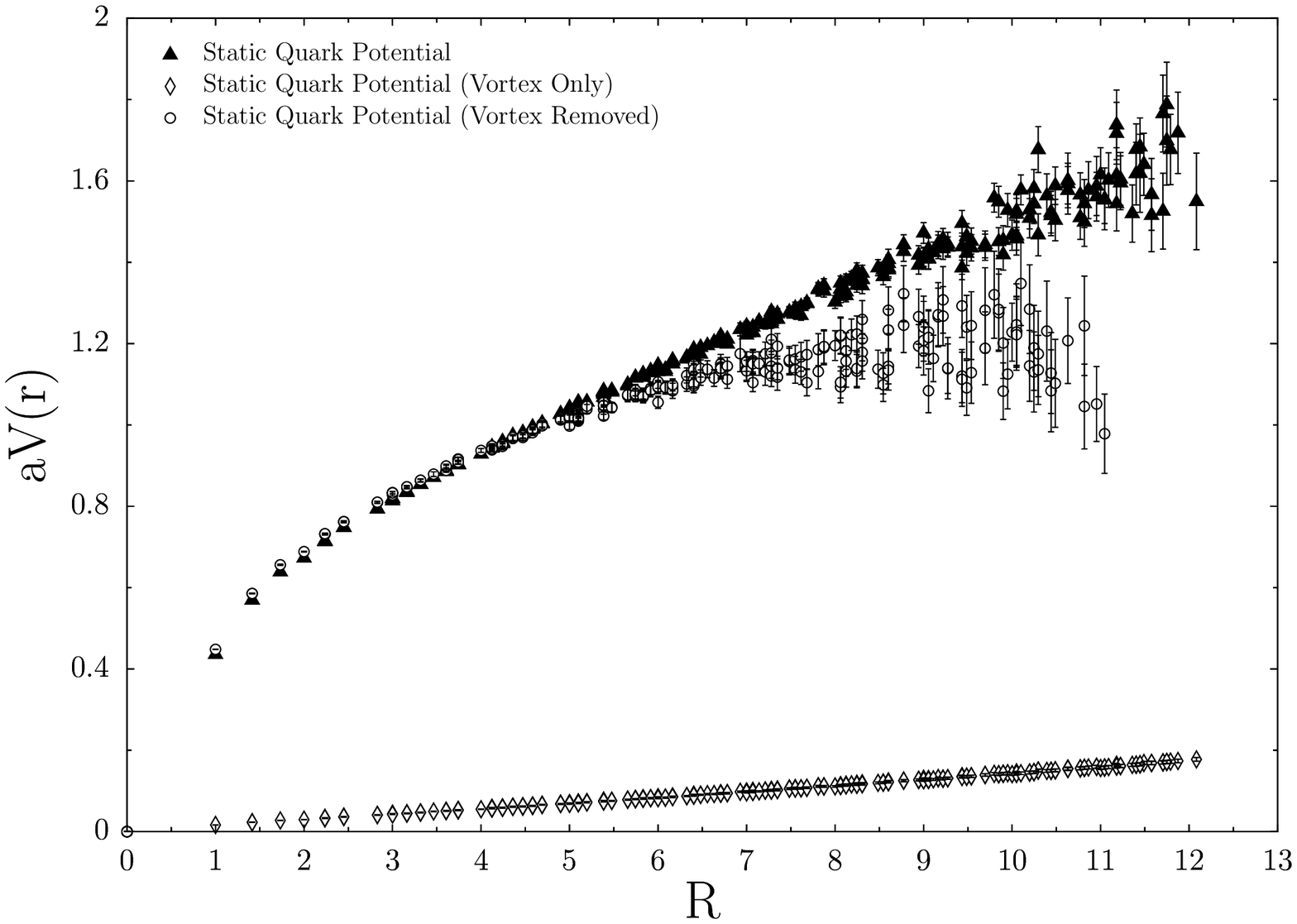}
  
\end{array}$
\caption{Static quark anti-quark potential plots for each of preconditioning smearing sweeps used. Each plot contains data for the full, vortex-removed and vortex-only configurations.The data shown uses a 3 timeslice fit window in each case with the fit window being from timeslice 4 to 6 for the full data, timeslice 5 to 7 for the vortex-removed data and timeslice 10 to 12 for the vortex-only data.}
\label{figSQPs}
\end{figure*}

\begin{figure*}[h]
\hspace{-1cm}
$\begin{array}{c@{\hspace{1cm}}c}
\mbox{\bf No Smearing Preconditioning} & \mbox{\bf 80 Sweeps OverImp. Preconditioning}\\ 
\includegraphics[height=5.4cm]{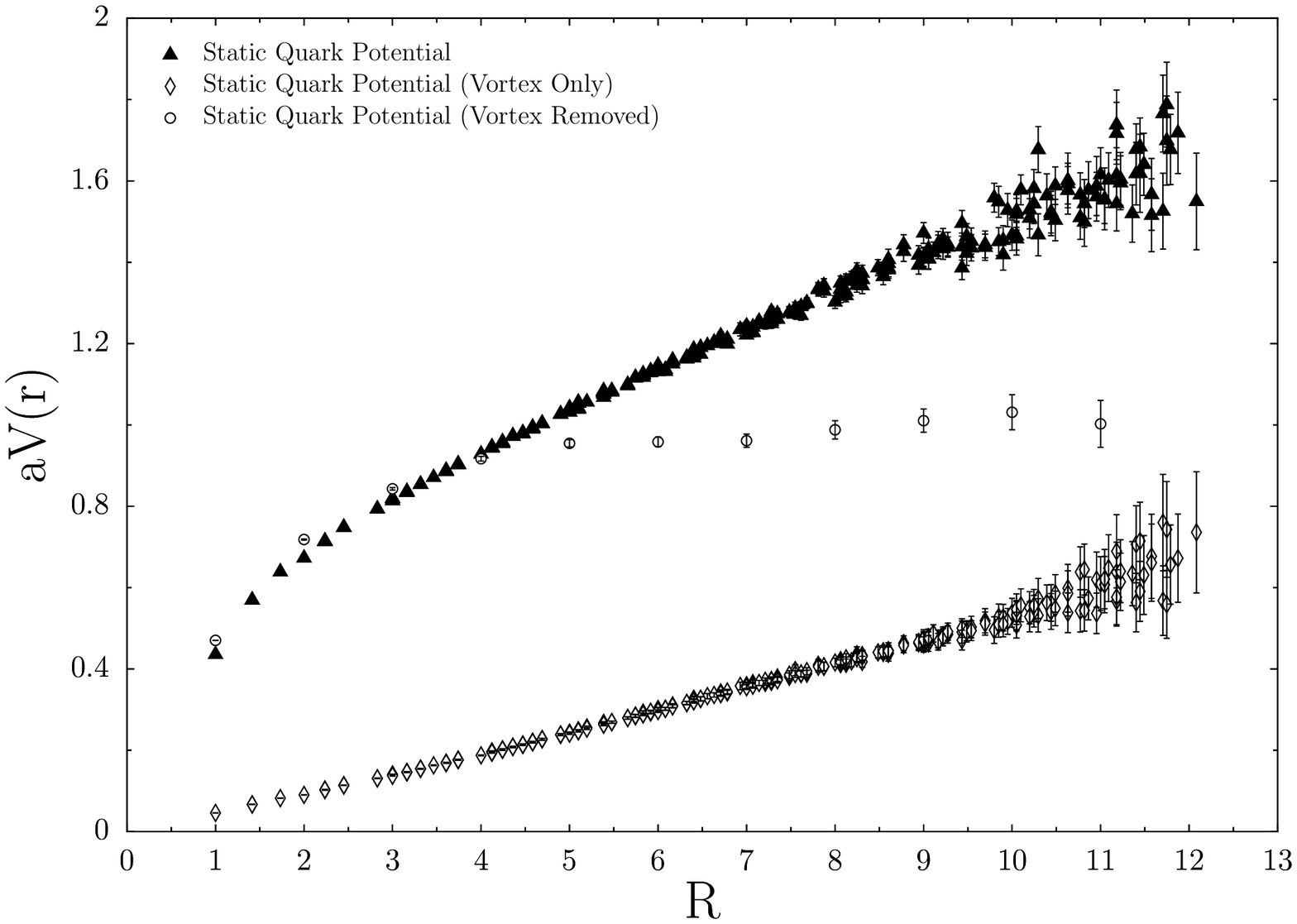} &
        \includegraphics[height=5.4cm]{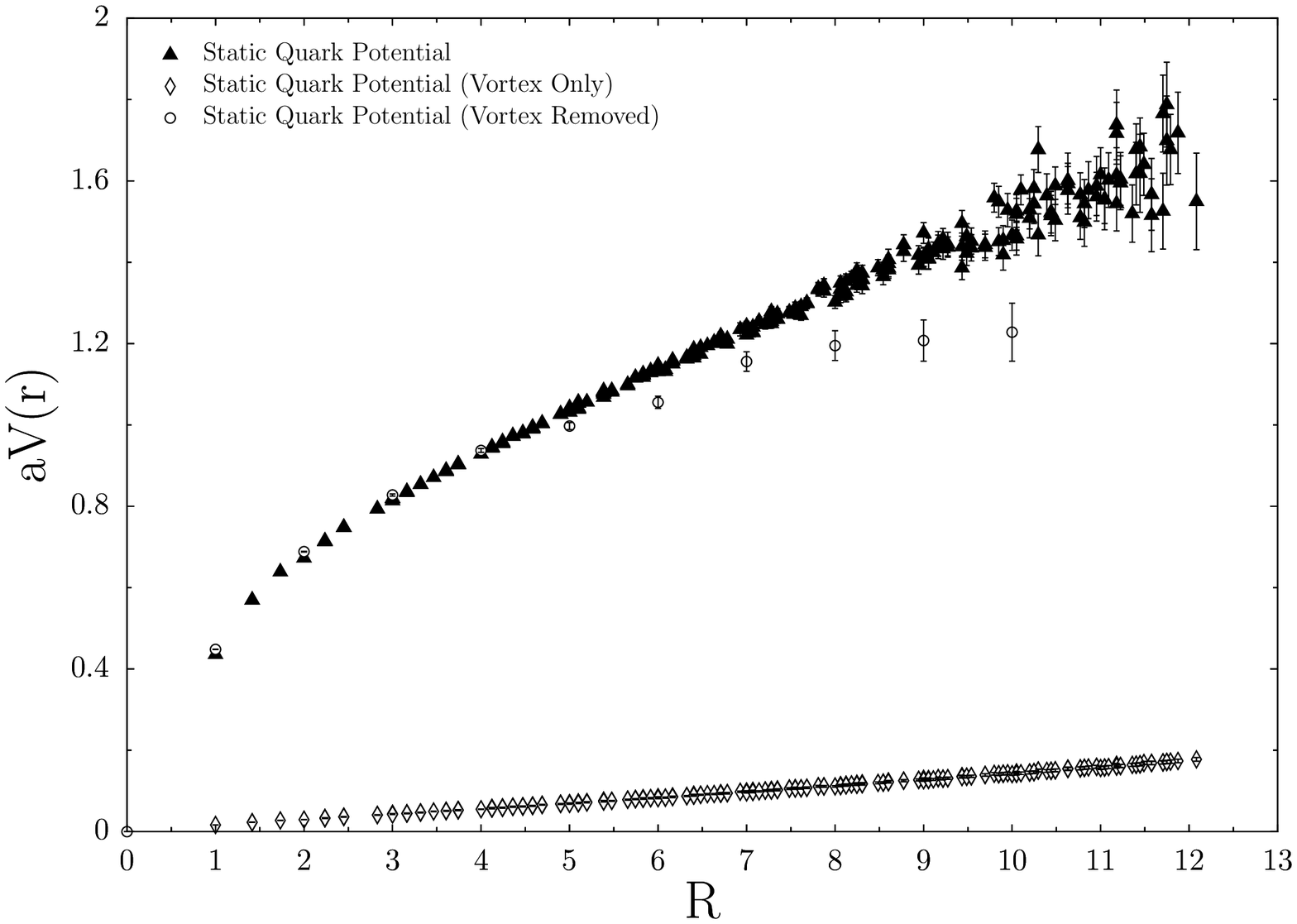} 

\end{array}$
\caption{The static quark anti-quark potential plots for both the lowest (left) and highest (right) levels of preconditioning. Only the on-axis data is shown for the vortex-removed configurations.}
\label{figSQPOnAxis}
\end{figure*}

Of significant concern when comparing the potentials of the unpreconditioned and preconditioned results in Fig.~\ref{figEffms} is the direct comparison of the potentials in each case. The top figures show the potential after gauge-fixing but prior to the center projection and vortex-removal and, of course, since the static quark potential is gauge invariant these plots are identical. For the lowest plots, which contain the vortex-removed data, we can see that the rate of decay for the preconditioned data has dropped and the quality of the data does not allow us to see whether it is possible that it plateaus at the same level as the unpreconditioned data. Of course the most dramatic effect occurs in the middle plots with the vortex-only data. There is a dramatic reduction in the magnitude of potential for all separations and this is direct manifestation of the Gribov-copy effect for this gauge-fixing method.

This Gribov-copy effect is also manifest in Fig.~\ref{figSQPs}. Here we show plots of the effective potential as a function of separation for each of the six levels of preconditioning used as well as the unpreconditioned data. It would appear that the findings are consistent with loss of confinement upon P-vortex removal. Although it would seem that this is perhaps not such a reasonable observation in the over-improved case, if we look exclusively at the on-axis contributions to the potential in this case (Fig.~\ref{figSQPOnAxis}) and compare it to that of the unpreconditioned case, we observe that a plateau in the potential may exist but at larger values of the separation. This would concur with our previous observation that the potential takes longer to plateau in this case and therefore the fit window may not be adequately account for this effect. Careful examination of the vortex-removed plot also reveals that we obtain an increasingly more accurate fit to the short-range Coulombic portion of the potential.

What is more significant however is that the value of the string-tension determined from the vortex-only configurations drops dramatically, and systematically, from $\sim60\%$ to as low as  $\sim16\%$ of the full string tension with increased preconditioning providing improvement. This is a disturbing manifestation of the Gribov problem since it perhaps questions how accurately we have determined the center vortices by our projection of the P-vortices with our fixing method.

\subsection{Discussion}

\begin{figure}
\hspace{-1.7cm}\includegraphics[height=7.1cm]{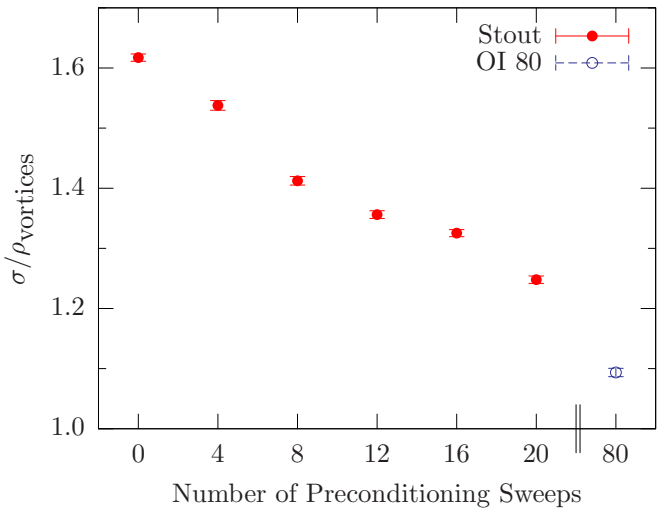} 
\caption{Ratio between the vortex-only string tension and the vortex density as a function of preconditioning.}
\label{figRatio}
\end{figure}

The use of smearing as a preconditioning technique does indeed lead to higher maxima in the MCG gauge-fixing condition $V_U[\Omega]$. These higher maxima in turn lead to lower numbers of P-vortices determined in the center projection. In $SU(2)$, similar results have been obtained when seeking higher maxima by means of simulated annealing \cite{Bornyakov:2000ig} and by pre-fixing to Landau gauge prior to MCG fixing \cite{Kovacs:1999st}. As observed in $SU(2)$ \cite{Bornyakov:2000ig,Faber:2001hq}, there appears to be a significant anti-correlation between the value achieved in the gauge fixing functional and the percentage string tension reproduced by center vortices. 

As can be seen in Fig.~\ref{figRatio}, the ratio between the vortex-only string tension and the vortex density (simply the fraction of vortex plaquettes to total number of plaquettes) as a function of preconditioning is not independent of the preconditioning. Had it been independent one might conclude the reduction in the string tension is associated with simply not identifying all the vortices present. Either the mechanism with which vortices produce confinement is not entirely intact or the physical relevance of the vortices is not uniformly distributed. 

In $SU(2)$, it was seen that smearing an $SU(2)$ configuration prior to MCG fixing reduced the center projected string tension considerably \cite{Del Debbio:1998uu}. It was argued there that this is because smearing greatly expands the vortex cores making the MCG process of collapsing them to pierce a single plaquette more difficult. A similar point was used to address the issue raised by prefixing to Landau gauge \cite{Kovacs:1999st}. In principle the same position could be taken here, the generated preconditioning transformation may allow the vortex cores to be distributed across a larger number of lattice sites and again make the MCG task of compressing them more difficult. However, the over-improved stout-link smearing parameters are deliberately chosen to maintain the size of instantons and there is a case to say that if the link between center vortices and topology seen in $SU(2)$ persists in $SU(3)$ then it should be possible to smear configurations without expanding the vortex cores. It is difficult to attribute the same vortex-expanding behavior to the case of simulated annealing. However, the fact that known higher
maxima exist (having these properties) and that simulated annealing is
designed to locate them could explain the similar behavior.

It is significant that much of the discussion in Ref.~\cite{deForcrand:2000pg}, where there is no Gribov ambiguity, can also be reconciled with the results found here. In this case a number of different Laplacian operators were constructed simply by using smeared links in the definition of the operator. There too it was seen that this caused an analogous effect on the vortex-only string tension. It was argued that the use of smearing caused the Laplacian to be blind to the short-range physics making the decomposition of the gauge field into the confining and non-confining components less effective --- disorder in the vortex-only component is absorbed into the vortex-removed component resulting in a loss of string tension. It was contended there that in the continuum limit the smearing radius shrinks to zero, restoring the string tension. 

In the same way, the smearing preconditioning may allow this effect to occur for MCG. That the locations of vortices as determined by both methods coincide serves to strengthen this position. Indeed, as discussed in Ref.~\cite{deForcrand:2000pg}, periodic boundary conditions cause gauge defects to have an opposite partner and, perhaps, the non-locality introduced by the preconditioning procedure may allow these opposites to annihilate producing no net defect after projection and a resultant drop in the vortex-only string tension. 

\section{Conclusions} 

The use of smearing as a preconditioning technique leads to higher maxima in the MCG gauge-fixing condition $V_U[\Omega]$. These higher maxima in turn lead to lower numbers of P-vortices determined in the center projection and, subsequently, lower values of the vortex-only string tension.

Although the fundamental modular region of MCG would be an ideal candidate for a unique definition of vortex texture, it seems that the vortex matter arising from the first Gribov region as a whole has a greater phenomenological relevance. While all preconditionings lead to a loss of string tension, it is the center-projected physics that is not consistent. An improvement in $V_U[\Omega]$ causes one to miss vortices in the projection and spoil the phenomenology. 

While MCG has proved successful to a large extent in $SU(2)$, what is different in $SU(3)$ is that  center-projection has never been shown to find enough vortices to reproduce the full string tension \footnote{Apart from possibly vortices as determined via Laplacian gauge \cite{deForcrand:2000pg} in the continuum limit.}. It just may be that MCG gauge fixing criterion is not sufficient to accurately locate the center vortices in $SU(3)$ and that reproducing the success in $SU(2)$ is a matter of getting closer to an ``ideal'' gauge. Nevertheless, it would be informative to look for correlations between the locations of the determined P-vortices with each preconditioning since their removal still leads to a loss of string tension.

\end{document}